\documentclass[12pt]{article}
\pdfoutput=1
\usepackage{epsf,amsfonts,amssymb,epsfig,amsmath,mathtools,graphics,slashed}
\usepackage{hyperref,hep,graphicx,subfig}
\usepackage{rotating}
\usepackage[final]{showlabels}
\allowdisplaybreaks

\makeatletter

\@addtoreset{equation}{section}
\makeatother

\begin{document}

\begin{titlepage}

\vfill

\begin{flushright}
%Imperial/TP/201?/JG/??\\
DCPT-19/03
\end{flushright}

\vfill

\begin{center}
   \baselineskip=16pt
   {\Large\bf Holographic transport and density waves}
  \vskip 1.5cm
  \vskip 1.5cm
      Aristomenis Donos and Christiana Pantelidou\\
   \vskip .6cm
      \begin{small}
      \textit{Centre for Particle Theory and Department of Mathematical Sciences,\\ Durham University,
       Durham, DH1 3LE, U.K.}
        \end{small}\\   
%    \vskip .6cm
%      \begin{small}
%      \textit{$^1$ Centre for Particle Theory and Department of Mathematical Sciences,\\ Durham University,
%       Durham, DH1 3LE, U.K.}
%        \end{small}\\
%      \vskip .6cm
%      \begin{small}
%      \textit{$^2$Blackett Laboratory, 
%        Imperial College\\ London, SW7 2AZ, U.K.}
%        \end{small}\\ *[.6cm]
         
\end{center}

\vfill

\begin{center}
\textbf{Abstract}
\end{center}
\begin{quote}
We consider transport of heat and charge in holographic lattices which are phases of strongly coupled matter in which translations are broken explicitly. In these systems, we study a spontaneous density wave that breaks translations incommensurately to the lattice. The emergent gapless mode due to symmetry breaking couples to the heat current impacting transport at low frequencies. We study the effects of this coupling when the mode is freely sliding as well as after the introduction of a small deformation parameter which pins down the density wave. We prove that the DC transport coefficients are discontinuous in the limit of the pinning parameter going to zero. From the perspective of finite frequency thermoelectric conductivity, this limiting process is accompanied by the transfer of spectral weight to frequencies set by the pinning parameter. As expected, for weak momentum relaxation, this spectral weight transfer appears as a shift of the Drude peak.
\end{quote}

\vfill

\end{titlepage}

\setcounter{equation}{0}

\section{Introduction}
Holography provides a framework for studying large classes of strongly coupled field theories. One of its most attractive aspects is the powerful  techniques of classical gravity that one can use in the bulk in order to study arbitrary deformations and sources of the dual field theory.

With a view to applications in condensed matter systems \cite{Herzog:2009xv,Hartnoll:2009sz,Hartnoll:2016apf}, we are particularly interested in field theories which have a global $U(1)$ and have been deformed by the corresponding chemical potential. Another set of deformations that we wish to consider break spatial translations explicitly and thus eliminate momentum from the conserved charges of the system. This is necessary in order to obtain finite low frequency response of heat and charge currents upon application of an external electric field and temperature gradient. The resulting black hole spacetimes are known as holographic lattices \cite{Hartnoll:2012rj} and are rather complicated as one needs to solve a coupled system of non-linear PDEs in order to explicitly construct them \cite{Horowitz:2012ky,Chesler:2013qla,Donos:2014yya}.

In order to simplify the problem, we will consider holographic lattices which lead to homogeneous bulk spacetimes such as the helical lattices \cite{Donos:2012js}, the Q-lattices \cite{Donos:2013eha} or the linear axion models of \cite{Andrade:2013gsa}. In this paper we have chosen to use a Q-lattice construction which allows us to consider isotropic lattice backgrounds. We will study CFT's with two spatial dimensions. Explicit breaking of translations in both spatial directions via Q-lattices requires two global $U(1)$'s in the bulk. The technical advantage of these constructions is that due to homogeneity, they lead to ODEs which are much simpler. Even though we will use this special class of lattices, we expect that our main results are general.

Holography also provides a powerful tool to study phase transitions and spontaneous symmetry breaking via black hole instabilities. Since the original introduction of holographic superfluids \cite{Gubser:2008px,Hartnoll:2008kx}, there has been a lot of progress in understanding different order parameters and corresponding phases of holographic matter. For the purposes of our paper, we will be interested in black hole instabilities that lead to spontaneous breaking of translations. Such instabilities lead to helical phases \cite{Nakamura:2009tf,Donos:2012wi}, and more generally to density waves \cite{Donos:2011bh, Donos:2013gda} of different patterns \cite{Withers:2014sja,Donos:2015eew}.

In this paper we are interested in the physics of the sliding mode associated with the density wave\footnote{In order to avoid confusion we note that in this paper we will study systems in which a scalar operator develops a VEV with a non-trivial density profile rather than the time component of a conserved current density.}. The existence of a gapless sliding mode for the density waves requires the breaking of a continuous bulk symmetry in addition to translations which is already broken explicitly by the holographic lattice. This suggests that we should consider two additional global $U(1)$'s in the bulk that will precisely play this role. As we will see, we can mathematically construct black holes solutions in which the order parameter associated to the breaking of these additional $U(1)$'s will also break translations. In the models we will consider, these solutions are going to be thermodynamically subdominant to solutions in which the order parameter remains translationally invariant. One can think of more complex holographic models in which a translationally non-invariant order parameter will be thermodynamically preferred. However, we expect that the point we are making here will remain in even more complex holographic models in which a translationally non-invariant order parameter will be thermodynamically preferred. The novelty of our system is that it contains a sliding mode despite the explicit breaking of translations. This allows us to introduce a pinning parameter which can be chosen to be completely independent from the other scales in the system, including the strength of the holographic lattice. This is in contrast with the systems used previously when considering the effects of weakly pinning a density wave\footnote{The case of holographic transport in the presence of spontaneous density waves without any explicit breaking of translations has been studied in \cite{Amoretti:2017frz,Amoretti:2017axe,Donos:2018kkm,Gouteraux:2018wfe}.} in the context of holography  \cite{Andrade:2017ghg,Andrade:2017leb,Andrade:2017cnc,Andrade:2015iyf,Amoretti:2018tzw,Andrade:2018gqk} \footnote{For a more phenomenological approach to spontaneous breaking of translations and pinning of density waves see \cite{Alberte:2017oqx,Alberte:2017cch}}.

From the point of view of the boundary theory, the coexistence of an order parameter that breaks translations with a background lattice is of interest in condensed matter systems which exhibit density wave instabilities incommensurate to the atomic lattice. As long as translations are broken explicitly and the $U(1)$ carrying the chemical potential remains unbroken, the electric conductivity of the system will remain finite. However, the sliding mode resulting from symmetry breaking will be gapless and will also couple to the heat current leaving a significant impact on transport at low frequencies. The aim of our paper is to study precisely the effects of this coupling in the context of holographic theories.

As we will show, the addition of a small deformation parameter $\phi_{s}$ that pins down the sliding mode introduces a drastic change in the DC transport properties. Within holography, DC transport in the absence of such sliding modes has been extensively studied \cite{Policastro:2001yc,Iqbal:2008by,Blake:2013bqa,Donos:2014cya,Donos:2015gia,Banks:2015wha}]. It has been shown that DC transport coefficients are completely fixed after solving for  a Stokes flow of a charged fluid living on the curved horizon of a black hole \cite{Donos:2015gia,Banks:2015wha}. The end product of this procedure gives the DC transport coefficients in terms of black hole horizon data and thermodynamics. When the density wave is pinned down, the previous results in the literature are applicable and can be directly imported to the model we will consider here. However, when the density wave is not pinned, the sliding mode needs to be taken into account discontinuously modifying the transport coefficients as functions of the pinning parameter. The effects of pinning of density waves have been appreciated before in the context of condensed matter physics \cite{Gruner}.

As we show explicitly in the main text, the diagonal components of the matrix of transport coefficients with pinning are smaller than in the freely sliding case. The physics of this DC transport discontinuity becomes more transparent when seen from the point of view of transport at finite frequency. Away from $\omega=0$, the optical conductivity is going to be a smooth function of $\phi_{s}$. The discontinuity at $\omega=0$, $\phi_{s}=0$ is a result of trying to commute the limits $\phi_{s}\to 0 $ and $\omega\to 0$. The sum rules dictate that the total spectral weight in the conductivities is going to be only perturbatively affected when varying $\phi_{s}$ away from zero. This suggests that when the density wave is even weakly pinned, spectral weight from the origin is going to be transferred up to frequencies whose scale is determined by $\phi_{s}$. In the case of weak momentum relaxation rates, this effect appears as a shift of the Drude away from the origin.

Our paper is structured in five sections. Section \ref{sec:background} introduces the class of models that we will study and discusses the numerical construction of the background black holes. Section \ref{sec:thermodynamics} is devoted to holographic renormalisation and thermodynamics. In section \ref{sec:DC}, we will study the DC transport properties of the thermal states and we will express the thermoelectric conductivities in terms of horizon data and thermodynamic quantities. In section \ref{sec:AC}, we will study the problem of AC transport numerically. Firstly, we will confirm that the analytic formulae for DC transport formulae of section \ref{sec:DC} are indeed equal to the zero frequency limit of the finite frequency conductivities. Secondly, we will introduce a small pinning parameter to the system in order to demonstrate the controlled shift of the Drude peak away from zero frequency. We conclude our paper with a discussion in section \ref{sec:discussion} where we also point at different projects of interest to us.

\section{Set-up and background solutions}
\label{sec:background}
We consider the following four-dimensional Einstein-Maxwell action coupled to six real scalars, $\phi$, $\psi$, $\chi_i$ and $\sigma_i$ with $i=1,2$,
\begin{multline}
\label{eq:action}
S=\int d^4x \sqrt{-g}\left(R-V(\phi,\psi)-\frac{3}{2}(\partial\phi)^2-\frac{1}{2}(\partial\psi)^2-\frac{\tau(\phi,\psi)}{4}F^2\right.\\
\left.-\frac{\theta(\phi)}{2}[(\partial \chi_1)^2+(\partial \chi_2)^2]-\frac{\theta_1(\psi)}{2}[(\partial \sigma_1)^2+(\partial \sigma_2)^2]\right)\,.
\end{multline}
The global shift symmetries of $\chi_{i}$ and $\sigma_{i}$ make this theory suitable for Q-lattice constructions \cite{Donos:2013eha}. Such constructions have been extensively used in the past to explicitly break spatial translations from the boundary point of view. In order to make contact with the Q-lattice models, we can think of  \eqref{eq:action} as a consistent truncation of a theory that contains four scalars $Z_{j}$ and $W_{j}$ taking values in a complex target space with $Z_{j}=\rho(\phi_{j})\,e^{i\,\chi_{j}}$ and $W_{j}=R(\psi_{j})\,e^{i\,\sigma_{j}}$. All the solutions of the model \eqref{eq:action} we will consider in this paper can be seen as solutions of the complex scalar model with $\phi_{j}=\phi$ and $\psi_{j}=\psi$.

As we will explain later in more detail, by choosing appropriate boundary conditions we will use the scalars $\psi$ and $\sigma_{i}$  to break translations explicitly with a period of wavenumber $k_{s}$ while keeping the system at finite temperature and chemical potential. The physical motivation behind this choice is to relax momentum in the system and study its low frequency transport properties.

The scalar $\phi$ will correspond to a relevant operator of the UV theory. In the normal phase black holes $\phi$ is going to be trivial everywhere in the bulk. Depending on the details of the functions $V$, $\tau$ and $\theta$, below a certain temperature the field $\phi$ can develop instabilities which will lead to a broken phase branch of black holes. In our language, the scalars $\phi$ and $\chi_{i}$ will be used to study the density wave of wavelength $k$. Since the breaking spontaneous, the dual operators will not carry any deformation.

Thinking of constructing these broken phase solutions, the scalars $\chi_{i}$ can be taken to depend linearly on the spatial coordinates of the boundary theory with the constant of proportionality $k$ playing the role of a wavelength. Note that $k$ can in general be different from $k_{s}$ leading to the order parameter breaking translations incommensurately to the background lattice. Moreover, one can add a constant to each of the scalars $\chi_{i}$ and still satisfy the boundary conditions which we will choose to be consistent with spontaneous symmetry breaking. This is precisely a Goldstone mode in the bulk which will give rise to a massless mode from the boundary theory point of view. When $k$ is non-zero, this massless mode will couple to the heat current.

The black hole solution ansatz that captures the above ingredients is given by 
\begin{align}
\label{eq:ansatz}
&ds^2=-U(r) dt^2+\frac{1}{U(r)} dr^2+ e^{2 V_1(r)} \delta_{ij} dx^i dx^j\,,\nonumber\\
&A=\alpha(r) dt\,,\quad \phi=\phi(r)\,,\quad \chi_i= k x^i\,,\quad\psi=\psi(r)\,,\quad \sigma_i= k_s x^i\,,
\end{align}
where $i=1,2$ and $k,k_s$ are the wavelengths of the spontaneous and explicit breaking respectively. The variation of the action \eqref{eq:action} gives rise to the following field equations of motion
\begin{align}
\label{eq:eom}
&R_{\mu\nu}-\frac{\tau}{2} (F_{\mu\rho}F_{\nu}{}^{\rho}-\frac{1}{4}g_{\mu\nu}F^2)-\frac{1}{2}g_{\mu\nu} V-\frac{3}{2}\partial_\mu\phi\partial_\nu\phi-\frac{1}{2}\partial_\mu\psi\partial_\nu\psi\notag\\
&\quad-\sum_{i}(\frac{\theta}{2}\partial_\mu\chi_i\partial_\nu\chi_i+\frac{\theta_1}{2}\partial_\mu\sigma_i\partial_\nu\sigma_i)=0\,,\nonumber\\
& \frac{3}{\sqrt{-g}}\partial_\mu\left(\sqrt{-g}\,\partial^\mu\phi\right)-\partial_{\phi}V-\frac{1}{4}\partial_{\phi}\tau\, F^{2}-\frac{1}{2}\theta^{\prime}\,\sum_{i}(\partial \chi_i)^{2}=0\,,\nonumber\\
& \frac{1}{\sqrt{-g}}\partial_\mu\left(\sqrt{-g}\,\partial^\mu\psi\right)-\partial_{\psi}V-\frac{1}{4}\partial_{\psi}\tau\, F^{2}
-\frac{1}{2}\theta'_{1}\,\sum_{i}(\partial \sigma_i)^{2}=0\,,\nonumber\\
& \frac{1}{\sqrt{-g}}\partial_\mu\left(\theta_1\sqrt{-g}\,\partial^\mu\sigma_i\right)=0\,,\quad \frac{1}{\sqrt{-g}}\partial_\mu\left(\theta\sqrt{-g}\,\partial^\mu\chi_i\right)=0\,,\nonumber\\
& \partial_\mu(\sqrt{-g}\, \tau F^{\mu \nu})=0\,.
\end{align}
For definiteness we will consider
\begin{align}
&V(\phi, \psi)=-6 \cosh \phi\,,\nonumber\\
&\theta(\phi)=12 \sinh^2(\delta\, \phi)\,,\nonumber\\
&\theta_1(\psi)=\psi^2\,,\nonumber\\
&\tau(\phi, \psi)=\cosh(\gamma\, \phi)\,,
\end{align}
but unless stated otherwise the main points of our paper are independent of this particular choice.  

In order to construct the backgrounds and eventually study AC transport, we will solve numerically the equations \eqref{eq:eom} to construct black hole solutions with temperature $T$ that asymptote to $AdS_4$ in the UV and are regular in the IR. The ansatz \eqref{eq:ansatz} provides a solution of the equations of motion \eqref{eq:eom} provided the functions appearing in \eqref{eq:ansatz} satisfy a system of four second order and one first order ODEs. This counting implies that the solution to this system of equations is fully determined by nine constants of integration. The metric ansatz \eqref{eq:ansatz} is invariant under translations of the radial coordinate $r$ and we will use this freedom to set the horizon at $r=0$. To ensure that we have a regular Killing horizon we assume that we have the following expansions at $r = 0$
\begin{align}
\label{eq:IRexp}
U= 4\pi T\,r+\dots\,,\quad V_1=V_1^{(0)}+\dots\,,\quad \alpha=\alpha^{(0)}\,r+\dots\,,\nonumber\\
\phi=\phi ^{(0)}+\dots\,,\quad\psi=\psi^{(0)}+\dots\,.
\end{align}
where we have four constants of integration to be fixed via a double sided shooting method.

In the UV, we have the following expansion
\begin{align}
\label{eq:UVexp}
U&= (r+R)^2+\dots+W \,(r+R)^{-1}+\dots\,,\nonumber\\
V_1&=\log(r+R)+\dots\,,\nonumber\\
\alpha&=\mu+Q \,(r+R)^{-1}+\dots\,,\nonumber\\
\phi&=\phi_{s}\, (r+R)^{-1}+S_V\, (r+R)^{-2}+\dots\,,\nonumber\\
\psi&=\psi_s+\dots+\psi_V\, (r+R)^{-3}+\dots\,.
\end{align}
The constants of integration $\mu$, $\phi_{s}$ and $\psi_{s}$ represent sources of the dual field theory and are going to be held fixed. In particular $\mu$ will be the chemical potential of the theory and $\psi_{s}$ is the source for the explicit lattice we will consider. On the other hand $\phi_{s}$ will be set to zero for most of our calculatons in order for the asymptotics of $\phi$ and $\chi_{i}$ to be consistent with spontaneous symmetry breaking. Given these boundary conditions, the constants $R$, $W$, $Q$, $S_{V}$ and $\psi_{V}$ along with the four constants from the near horizon expansion \eqref{eq:IRexp} yield the total of nine necessary constants. Using a shooting technique we will construct black hole solutions with $\{k/\mu,k_s/\mu,\phi_{s}/\mu,\psi_s,\gamma,\delta\}=(\tfrac{15}{100},\tfrac{3}{10},0,4,3,\tfrac{1}{2})$ and various values of $T/\mu$.

In order to understand the phase diagram of our theory, it is useful to point out that for no lattice deformation $\psi_{s}=0$, the unbroken phase black hole solution is the electrically charged Reissner-N\"ordstrom black hole with
\begin{align}\label{eq:RN}
&U= (r+R)^{2}-\left(R^{2}+\frac{\mu^{2}}{4} \right)\frac{R}{r+R}+\frac{\mu^{2}R^{2}}{4\,(r+R)^{2}},\quad V_{1}=\ln (r+R),\notag\\
&a=\mu\,\frac{r}{r+R},\quad \phi=0,\quad \psi=0,\notag\\
&R=\frac{2}{3}\,\pi T+\frac{1}{6}\,\sqrt{16\pi^{2}T^{2}+3\mu^{2}}\,.
\end{align}

The near horizon limit of the $T=0$ RN black hole is given by the $AdS_{2}\times R^{2}$ solution with
\begin{align}\label{eq:AdS2}
U=6\,r^{2}, \quad V_{1}=V_{1}^{(0)},\quad a=2\sqrt{3}\,r,\quad \phi=0, \quad \psi=0\, ,
\end{align}
where $V_{1}^{(0)}=\ln\,(\mu/\sqrt{12})$. It is now enlightening to consider the static modes $\delta \phi=\varepsilon\,r^{\delta_{\phi}^{(\pm)}}$ and $\delta \psi=\varepsilon\,r^{\delta_{\psi}^{(\pm)}}$
of the scalars around the ground state \eqref{eq:AdS2}\,. Plugging in this perturbation in the equations of motion and expanding in $\varepsilon$ we find that
\begin{align}\label{eq:exps}
\delta_{\psi}^{(\pm)}&=-\frac{1}{2}\pm \frac{1}{6}\,\sqrt{9+12\,e^{-2V_{1}^{(0)}}\,k_{s}^{2}}\notag\\
\delta_{\phi}^{(\pm)}&=-\frac{1}{2}\pm\frac{1}{6}\,\sqrt{-3(1+4\gamma^{2})+48e^{-2V_{1}^{(0)}}\delta^{2}\,k^{2}}\,.
\end{align}
The above expressions reveal that the explicit lattice associated to $\psi$ and $\sigma_{i}$ corresponds to an irrelevant operator of the IR solution \eqref{eq:AdS2}. As a result, we have that the near horizon limit of the zero temperature limit of our unbroken phase black holes will again be given by \eqref{eq:AdS2} even after introducing the explicit lattice.

\begin{figure}[h!]
\centering
\includegraphics[width=0.48\linewidth]{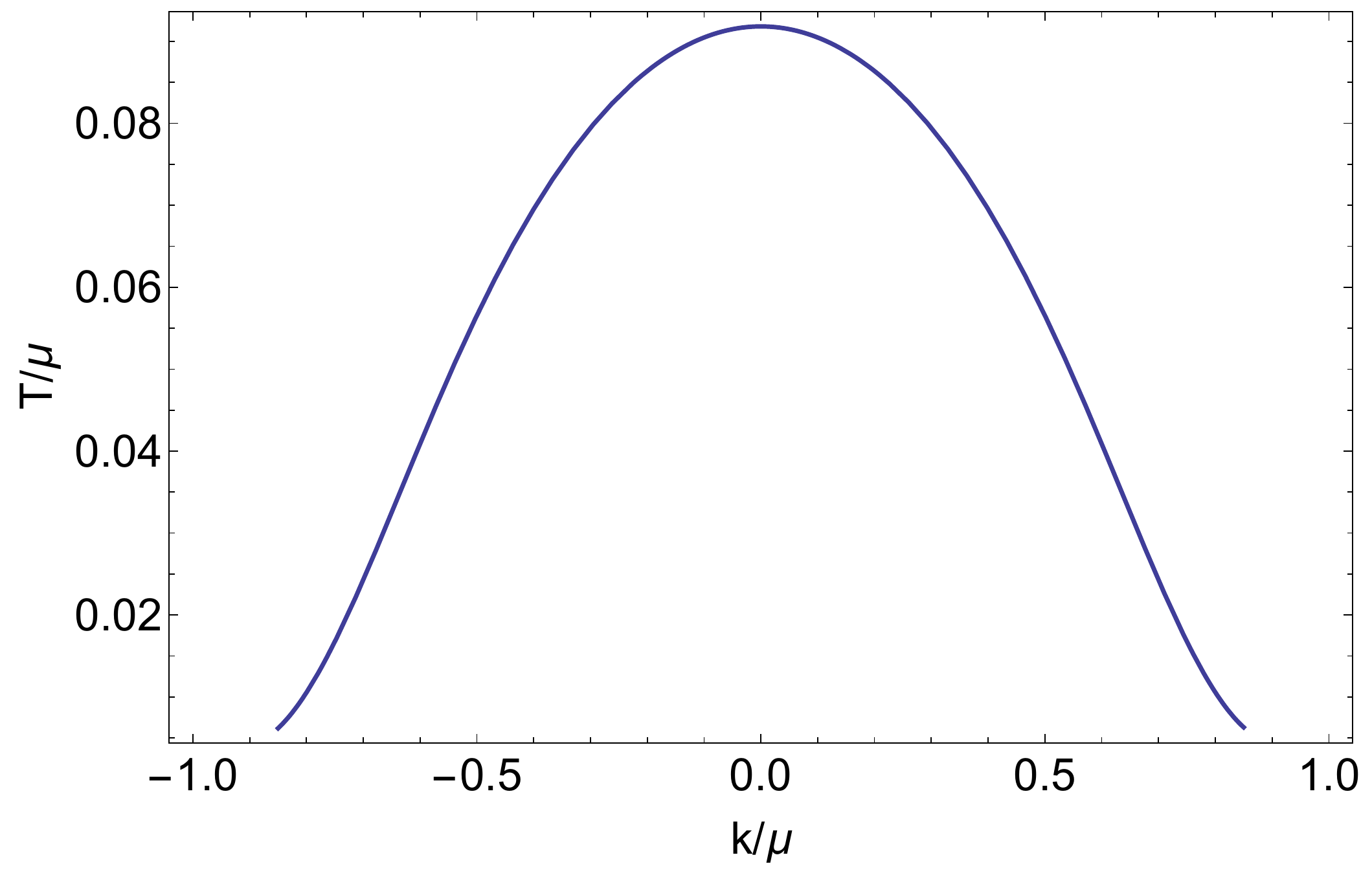}
\caption{Plot of the temperature $T$ at which a zero more for $\phi$ appears as a function of the wavelength $k$.}
\label{fig:BellCurve}
\end{figure}

In additon, one can also see that there is a range of $k$ for which the exponent $\delta_{\phi}^{(\pm)}$ becomes complex and therefore $\phi$ and $\chi_{i}$ yield a mode that violates the BF bound. This signifies the existence of a tachyonic mode that will make the previously described normal phase black holes unstable below a critical temperature which will depend on $k$. Exactly at the critical temperature there will be a zero mode for $\phi$ that will give rise to the new branch of broken phase black holes. From \eqref{eq:exps} we see that the lightest mode has $k=0$ and therefore we expect that this is going to be the broken phase black hole with the highest critical temperature. As a result, the corresponding branch will be the thermodynamically dominant black holes solution at finite temperature. As one increases the value of $k$, the corresponding critical temperature will decrease up to a maximum value of $k$ where it will go all the way down to zero. For the particular model we study, one can see a plot of the critical temperature $T_{c}$ as a function of $k$ in figure \ref{fig:BellCurve}.

In this paper, we will study the transport properties of those finite $k$ black holes. Even though they are not thermodynamically preferred in this model, the general ideas in our paper and derivation will remain valid even in models where the black holes that minimise the free energy will be at finite $k$.

\section{Thermodynamics}
\label{sec:thermodynamics}

In this section we will study the thermodynamics of the black holes constructed in section \ref{sec:background}. We can read off the area of the event horizon and since we are working in units with $ 16 \pi G=1$, we deduce that the entropy density is given by
\begin{equation}
s=4 \pi\,e^{2 V_1^{(0)}}\,.
\end{equation}

In order to analyse the thermodynamics we calculate the on-shell Euclidean action. To do this, we analytically continue the time coordinate $t=-i\tau$. We consider the total Euclidean action, $I_{tot}$, defined as
\begin{equation}
I_{\mathrm{Tot}}=I+I_{bdr}\,,
\end{equation}
where $I=-iS$ and $I_{bdr}$  is given by the following integral on the boundary $r\to \infty$:
\begin{equation}\label{ctermp}
I_{bdr}=\int dt dx dy\sqrt{\gamma}(-2 K+4+\frac{3}{2}\phi^2-\frac{1}{2}\theta_1 \nabla \sigma_I \cdot \nabla \sigma_I)
\end{equation}
Here $K$ is the trace of the extrinsic curvature of the boundary and $\gamma_{\mu\nu}$ is the induced boundary metric.
It is possible to write the bulk part of the Euclidean  on-shell action in a total derivative form
\begin{align}\label{act2waysp}
I_{OS}&=-\int dx^4(-e^{2 V_1} U' + e^{2 V_1} \cosh \gamma\phi \,a \,a')'\,,
%&=-\int dx^4 [(-2e^{2 V_1} U V_1'-\frac{1}{2} U e^{2 V_1} \psi \psi')'-(-k^2\theta+\frac{1}{2} U e^{2 V_1} \psi'^2)]
 \end{align}
We next define the free energy $W=T[I_{Tot}]_{OS}\equiv w Vol_3$.
Using the UV and the IR expansions we obtain the following expression for the free energy density:
\begin{align}
\label{eq:OSactionp}
w&= -2 W-3 S_V \phi_{s}+Q \mu-s T\,,
%&=W-\frac{3}{2}\psi_s\psi_v-\int dr (-k^2\, \theta+\frac{1}{2} U e^{2 V_1} \psi'^2)
\end{align}

The first law takes the form
\begin{align}
\delta w=&-s \delta T+Q \delta \mu-3 S_V \delta \phi_{s} -3\psi_v\,\delta\psi_s  \nonumber\\
&+\frac{\delta k}{k}\int dr \,2\, k^2 \, \theta
\end{align}

Given the renormalised theory, we can also compute the expectation value of the boundary stress-energy tensor. The relevant terms are given by
\begin{align}\label{stressy}
\langle \tilde{T}^{\mu\nu}\rangle\equiv\lim_{r\to\infty} &[-2K^{\mu\nu}+\gamma^{ \mu\nu}(2K-4- \frac{3}{2}\phi^2+\frac{1}{2}\,\theta_1 \nabla \sigma_I \cdot \nabla \sigma_I)-\theta_1 \nabla^{\mu} \sigma_I \nabla^{\nu} \sigma_I\nonumber\\
&+ 2(R^{\mu\nu}-\frac{1}{2}\gamma^{\mu\nu}R)]
\end{align}
Using the asymptotic expansion \eqref{eq:UVexp}, one obtains the boundary stress-energy tensor 
\begin{align}
T_{tt}=-2W-3 S_V \,\phi_{s}\, \quad T_{ii}=-W\,,
\end{align}
which is traceless for $\phi_{s}=0$.

\section{DC conductivity from horizon data}
\label{sec:DC}
In this section we will consider the DC transport properties of our broken phase black holes. We will significantly modify the original analysis of \cite{Donos:2014cya} by adding the contribution of the sliding mode which is present in the broken phase black holes. We will introduce a constant electric field $E$ and temperature gradient source $\zeta$ on the boundary. In order to achieve this, we consider the perturbation
\begin{align}\label{eq:pert_ansatz}
\delta A_{x_1}&= -E\,t+\zeta \,a(r) \,t +\delta \alpha_{x_1}(r)\,,\nonumber\\
\delta g_{ t x_1}&=-U(r)\,\zeta \, t+ \delta h_{ t x_1}(r)\,,\nonumber\\
\delta g_{ r x_1}&=\delta h_{ r x_1} (r)\,,\nonumber\\
\delta \chi_ 1&=\delta c_{g} \,t +\delta\chi_1(r)\,,\nonumber\\
\delta \sigma_ 1&=\delta\sigma_1(r)\,,
\end{align}
keeping terms up to leading order in the perturbations. Notice that we kept the metric perturbation $\delta g_{rx_{1}}$ which we could have set to zero by an appropriate choice of coordinates. However, here we choose a slightly different class of coordinate systems to perform our calculation and for which we will only need to specify the behaviour of the function $\delta h_{rx_{1}}$ close to the boundary and the horizon. In other words we will imagine that we are not solving for $\delta h_{rx_{1}}$ as long as we fix it in a way that satisfies the boundary conditions we will specify for it. The linear term in time that shows up in $\delta\chi_{1}$ is precisely capturing the bulk Goldstone mode that couples to the perturbation for the heat current.

The system of equations that we get after plugging the perturbation ansatz \eqref{eq:pert_ansatz} in the equations of motion \eqref{eq:eom} and expanding is an inhomogeneous system of second order of equations for $\delta a_{x_{1}}$, $\delta h_{t x_{1}}$, $\delta \chi_{1}$ and $\delta \sigma_{1}$ in addition to the constraint
\begin{align}\label{eq:cons}
\delta h_{rx_{1}}U\,e^{-2V_{1}}\left(\theta \,k^{2}+\theta_{1}\,k_{s}^{2} \right)+(E-a\,\zeta)\tau\,a^{\prime}+\zeta\,U^{\prime}-U\left( 2\zeta V_{1}^{\prime}+k_{s}\theta_{1}\delta\sigma_{1}^{\prime}+k\theta\,\delta\chi_{1}^{\prime}\right)=0\,.
\end{align}
As long as we satisfy the constraint \eqref{eq:cons} at a single value of $r$, the second order ODEs imply that this is going to be satisfied everywhere. We will later choose to impose the constraint in the near horizon limit.

To set up the problem, we examine the perturbation close to the conformal boundary to write the expansion
\begin{align}
\label{eq:DC_asexp}
\delta a_{x_{1}}&=\frac{\delta a_{x_{1}}^{v}}{r+R}+\frac{k_{s}^{2}\delta a_{x_{1}}^{v} \psi_{s}^{2}}{6\,(r+R)^{3}}+\mathcal{O}\left(\frac{1}{r^{4}}\right)\,,\notag\\
\delta h_{t x_{1}}&=\frac{\delta h_{t x_{1}}^{v}}{r+R}+\mathcal{O}\left(\frac{1}{r^{2}}\right)\,,\notag\\
\delta \chi_{1}&=\delta\chi_{1}^{v}+\frac{k\zeta}{6\,(r+R)^{2}}+\mathcal{O}\left(\frac{1}{r^{4}}\right)\,,\notag\\
\delta \sigma_{1}&=-\frac{k\zeta}{2\,(r+R)^{2}}+\frac{\delta\sigma_{1}^{v}}{(r+R)^{3}}+\mathcal{O}\left(\frac{1}{r^{4}}\right)\,,\notag\\
\delta h_{rx_{1}}&=\mathcal{O}\left(\frac{1}{r^{3}}\right)\,,
\end{align}
where without loss of generality we have assumed that $\theta_{1}(\psi)\to \psi_{s}^{2}$ close to the conformal boundary. The above asymptotics ensure that we are not introducing any additional boundary sources apart from $E$ and $\zeta$ in \eqref{eq:pert_ansatz}. From the boundary point of view, the terms which are not directly proportional to either $E$ or $\zeta$ in \eqref{eq:pert_ansatz} carry information only about the VEVs of the dual operators. For the current operators we are interested in, we have
\begin{align}
\delta J^{\infty}&=\lim_{r\to \infty}\sqrt{-g}  \tau(\phi,\psi) \, F^{x_1\, r} =\delta a_{x_{1}}^{v} \,,\nonumber\\
\delta Q^{\infty}& =-T^{x_1}{}_t- \mu\, \delta J^{\infty}=-3\,\delta h_{t x_{1}}^{v}-\mu\,\delta a_{x_{1}}^{v}\,.
\end{align}
Regularity at the horizon is then achieved by imposing the following near horizon boundary conditions
 \begin{align}\label{eq:DC_nhexp}
&\delta \alpha_{x_1}= -\frac{E}{4\pi T}\log\,r+\delta \alpha_{x_1}^{(0)}+\dots\,,\nonumber\\
&\delta h_{ t x_1}=-v+\delta h^{(0)}_{ t x_1}\,r+\dots\,,\nonumber\\
&\delta h_{ r x_1}= -\frac{v}{4\pi T\,r}+\dots\,,\nonumber\\
&\delta \chi_ 1=\frac{\delta c_{g}}{4 \pi T}\,\log\,r+\delta\chi_{1}^{(0)}+\dots\,,\nonumber\\
&\delta \sigma_ 1=\delta\sigma_{1}^{(0)}+\dots\,.
\end{align}
At this point it is useful to establish the uniqueness of the solution we are after. From the expansions \eqref{eq:DC_asexp} and \eqref{eq:DC_nhexp} we count a total of ten constants $\delta a_{x_{1}}^{v}$, $\delta h_{t x_{1}}^{v}$, $\delta\chi_{1}^{v}$, $\delta\sigma_{1}^{v}$,  $\delta \alpha_{x_1}^{(0)}$, $\delta h^{(0)}_{ t x_1}$, $\delta\chi_{1}^{(0)}$, $\delta\sigma_{1}^{(0)}$, $\delta c_{g}$ and $v$. However, we only need nine in order to fully specify a solution to the system of four second order ODEs and the constraint \eqref{eq:cons}. The extra constant comes from the fact that asymptotically we are free to shift $\chi_{1}$ by a constant and still produce a valid solution to our boundary value problem and therefore only the difference $\delta\chi_{1}^{(0)}-\delta\chi_{1}^{v}$ can be uniquely fixed. Given this, we have now identified the nine integration constants that uniquely fix our solution and in particular, we see that  one of these is $\delta c_g$ which was introduced in the linear in time term for $\delta \chi_{1}$ in \eqref{eq:pert_ansatz}. Without it, we wouldn't be able to find a valid solution to our boundary value problem.

We now define the bulk quantities
\begin{align}
\delta J&=\sqrt{-g}\, \tau(\phi,\psi) F^{x_1\, r}=-\tau(\phi,\psi) \left( U(r)\,\delta \alpha_{x_1}'+\alpha' \delta h_{ t x_1}\right)\,,\\
\delta Q&=-a J+U^2\,\left(\frac{\delta h_{ t x_1}}{U}\right)'
\end{align}
for which the equations of motion \eqref{eq:eom} imply their radial evolution according to
\begin{align}
\partial_r \delta J&=0\,,\\
\partial_r \delta Q&=-\delta c_{g}\,k \, \theta\,,
\end{align}
Upon integration from the horizon to infinity the above equations can be used to relate the boundary electric current  and heat current pertubations $\delta  J^{(\infty)}$ and $\delta Q^{(\infty)}$ to horizon quantities along with an integral over the bulk
\begin{align}\label{eq:hor_const}
\delta  J^{(\infty)}&=\tau^{(0)}\,\left( a^{(0)}\,v+E\right) \notag\\
\delta Q^{(\infty)}&=4\pi T\,v-\delta c_{g}\,\int_{0}^{\infty}k\,\theta\,dr\,.
\end{align}

We now consider the gravitational constraint \eqref{eq:cons} close to the horizon from which we obtain the relation
\begin{align}\label{eq:contraint}
-4\pi T \zeta-\tau^{(0)}a^{(0)}\,E+\theta^{(0)}k\,\left( e^{-2V_{1}^{(0)}}k\,v+\delta c_{g}\right)+\theta_{1}^{(0)}e^{-2V_{1}^{(0)}}k_{s}^{2}v=0\,.
\end{align}
From the equations of motion for $\chi_{1}$ and $\sigma_{1}$ in \eqref{eq:eom} we obtain
\begin{align}
\partial_{r}\left(\theta\,e^{2V_{1}}U\,\partial_{r}\delta\chi_{1}\right)-k\,\theta\,\zeta-k\partial_{r}\left( \theta U\delta h_{rx}\right)=&0\label{eq:gstone_const}\\
\partial_{r}\left(\theta_{1}\,e^{2V_{1}}U\,\partial_{r}\delta\sigma_{1}\right)-k_{s}\,\theta_{1}\,\zeta-k_{s}\partial_{r}\left( \theta_{1} U\delta h_{rx}\right)=&0\,.
\end{align}
At this point it is crucial to highlight the difference between the spontaneous breaking related to $\phi$ and $\chi_{1}$ versus the explicit related to $\psi$ and $\sigma_{1}$. After integrating the above equations from the horizon up to infinity, we see that the first terms receive contributions from different boundary terms
\begin{align}
\theta^{(0)}\,e^{2V_{1}^{(0)}}\,\delta c_{g}+k\,\theta^{(0)}\,v+\zeta\,k\,\int_{0}^{\infty}\theta\,dr=&0\\
3\,\psi_{s}^{2}\,\delta\sigma_{1}^{v}+k_{s}\,\theta_{1}^{(0)}\,v+\zeta\,k_{s}\,\lim_{y\to\infty}\left(-\psi_{s}^{2}\,y+\int_{0}^{y}\theta_{1}\,dr\right)=&0
\end{align}
Apart from fixing the constant $\delta c_{g}$, the above point once again highlights the fact that not having a source for $\phi$ forces us to introduce the constant $\delta c_{g}$ in the perturbation \eqref{eq:pert_ansatz}. In more physical terms, there is a sliding mode which couples to our perturbation. The equations \eqref{eq:contraint}, \eqref{eq:gstone_const} and \eqref{eq:hor_const} can now be used to fix the boundary currents in terms of the sources
\begin{align}
\delta  J^{(\infty)}&=\left(\tau^{(0)}+\frac{{\tau^{(0)}}^{2}{a^{(0)}}^{2}}{\theta_{1}^{(0)}k_{s}^{2}e^{-2V_{1}^{(0)}}} \right)\,E+\frac{\tau^{(0)}a^{(0)}}{\theta_{1}^{(0)}k_{s}^{2}e^{-2V_{1}^{(0)}}}\left( 4\pi T+k^{2}e^{-2V_{1}^{(0)}}\,w_{k}\right)\,\zeta\\
\delta Q^{(\infty)}&=\frac{\tau^{(0)}a^{(0)}}{\theta_{1}^{(0)}k_{s}^{2}e^{-2V_{1}^{(0)}}}\left( 4\pi T+k^{2}e^{-2V_{1}^{(0)}}\,w_{k}\right)\,E\\
&\qquad\qquad\qquad +\left[\frac{\left( 4\pi T+k^{2}e^{-2V_{1}^{(0)}}\,w_{k}\right)^{2}}{\theta_{1}^{(0)}k_{s}^{2}e^{-2V_{1}^{(0)}}}+\frac{k^{2}}{\theta^{(0)}}e^{-2V_{1}^{(0)}}\,w_{k}^{2} \right]\,\zeta\\
w_{k}&\equiv \int_{r_{h}}^{\infty}\theta\,dr =\frac{1}{2k}\partial_{k}w\,.
\end{align}
From these expressions we obtain the transport coefficients
%\begin{align}
%\label{eq:DC_coeffs}
%\sigma&=\tau^{(0)}+\frac{{\tau^{(0)}}^{2}{a^{(0)}}^{2}}{\theta_{1}^{(0)}k_{s}^{2}e^{-2V_{1}^{(0)}}}\notag\\
%T\,\alpha=T\,\bar{\alpha}&=\frac{\tau^{(0)}a^{(0)}}{\theta_{1}^{(0)}k_{s}^{2}e^{-2V_{1}^{(0)}}}\left( 4\pi T+k^{2}e^{-2V_{1}^{(0)}}\,w_{k}\right)\notag\\
%T\,\kappa&=\frac{\left( 4\pi T+k^{2}e^{-2V_{1}^{(0)}}\,w_{k}\right)^{2}}{\theta_{1}^{(0)}k_{s}^{2}e^{-2V_{1}^{(0)}}}+\frac{k^{2}}{\theta^{(0)}}e^{-2V_{1}^{(0)}}\,w_{k}^{2}\,.
%\end{align}
\begin{align}
\label{eq:DC_coeffs}
\sigma&=\tau^{(0)}+\frac{4\pi\rho^{2}}{\theta_{1}^{(0)}k_{s}^{2}\,s}\notag\\
T\,\alpha=T\,\bar{\alpha}&=\frac{4\pi\,\rho}{\theta_{1}^{(0)}k_{s}^{2}}\left( T+(4\pi)^{-1}\,k^{2}e^{-2V_{1}^{(0)}}\,w_{k}\right)\notag\\
T\,\kappa&=\frac{4\pi\,s}{\theta_{1}^{(0)}k_{s}^{2}}\left( T+(4\pi)^{-1}\,k^{2}e^{-2V_{1}^{(0)}}\,w_{k}\right)^{2}+\frac{k^{2}w_{k}^{2}}{e^{2V_{1}^{(0)}}\theta^{(0)}}\,.
\end{align}
where
\begin{align}
\rho=e^{2V_{1}^{(0)}}\tau^{(0)}a^{(0)}\,,\quad s=4\pi\,e^{2V_{1}^{(0)}}\,,
\end{align}
are the charge and entropy density of the field theory respectively.  These expressions are to be contrasted with the ones we would find from the analysis with $\phi_{s}\neq 0$
% \begin{align}
%\label{eq:DC_coeffs_pinning}
%\sigma&=\tau^{(0)}+\frac{{\tau^{(0)}}^{2}{a^{(0)}}^{2}}{(\theta_{1}^{(0)}k_{s}^{2}+\theta^{(0)}k^{2})\,e^{-2V_{1}^{(0)}}}\notag\\
%T\,\alpha=T\,\bar{\alpha}&=\frac{4\pi T\,\tau^{(0)}a^{(0)}}{(\theta_{1}^{(0)}k_{s}^{2}+\theta^{(0)}k^{2})\,e^{-2V_{1}^{(0)}}}\notag\\
%T\,\kappa&=\frac{\left( 4\pi T\right)^{2}}{(\theta_{1}^{(0)}k_{s}^{2}+\theta^{(0)}k^{2})\,e^{-2V_{1}^{(0)}}}\,.
%\end{align}
 \begin{align}
\label{eq:DC_coeffs_pinning}
\sigma&=\tau^{(0)}+\frac{4\pi\rho^{2}}{(\theta_{1}^{(0)}k_{s}^{2}+\theta^{(0)}k^{2})\,s}\notag\\
T\,\alpha=T\,\bar{\alpha}&=\frac{4\pi T\,\rho}{\theta_{1}^{(0)}k_{s}^{2}+\theta^{(0)}k^{2}}\notag\\
T\,\kappa&=\frac{4\pi s\,T^{2}}{\theta_{1}^{(0)}k_{s}^{2}+\theta^{(0)}k^{2}}\,.
\end{align}
In particular the expressions \eqref{eq:DC_coeffs_pinning} are also valid for perturbative values of $\phi_{s}$, where the latter is parametrically smaller than any other scale in the system. An important point to note is that the background fields, and therefore the horizon data, are continuous in the $\phi_{s}\to 0$ limit. By directly comparing \eqref{eq:DC_coeffs} and \eqref{eq:DC_coeffs_pinning}, it is then obvious that in this regime the values of the DC transport coefficients without the pinning, \eqref{eq:DC_coeffs}, are larger and thus the limit $\phi_{s}\to 0$ will not be smooth as long we are in the broken phase where $S_{V}\neq 0$ and $k\neq 0$. The resolution to this discontinuity comes from considering the AC transport properties. It is natural to expect that for small values of $\phi_{s}$ there will be a finite frequency $\omega\approx \phi_{s}$ at which the real parts of the transport coefficients will exhibit a maximum that will be well approximated by \eqref{eq:DC_coeffs} while the zero frequency limit $\omega\approx 0$ will be given by \eqref{eq:DC_coeffs_pinning}. As we take the deformation parameter $\phi_{s}$ to zero, that maximum at $\omega\approx \phi_{s}$ will become the actual DC limit.
 
In the next section, we will carry out the numerical computation for the AC conductivity and demonstrate the above points. We will also show the validity of our new formulae \eqref{eq:DC_coeffs} for the DC transport coefficients by taking the small frequency limit of our AC computation.
 
\section{AC conductivity}
\label{sec:AC}
In this section we will use numerical techniques to study the optical conductivity around our thermal states. In section \ref{sec:AC_massless} we will consider the case where the  density wave is freely sliding and in section \ref{sec:AC_pinning} we will examine transport after having introduced a pinning parameter. As we will explain in detail, the difference between spontaneous and explicit breaking has some technical implications associated to the appropriate fixing of boundary conditions for the bulk perturbation.

It has been argued that weak momentum relaxation and the interplay between the strength of the density wave pinning and temperature could have an interesting application in the physics of transport in bad metals \cite{Delacretaz:2016ivq}. In the example we study in this section, and in particular in subsection \ref{sec:AC_pinning}, the lack of a Drude peak in the thermal optical conductivity in figure \ref{fig:AC_cond} shows that momentum relaxation is not small for that particular state. Despite that, in figure \ref{fig:AC_cond_pinning} we show that a parametrically small pinning parameter transfers spectral weight in a discontinuous way from $\omega=0$ to finite frequencies of the order of $\phi_{s}$, as expected from our general analysis of DC transport in section \ref{sec:DC}.

\subsection{AC transport with gapless modes}\label{sec:AC_massless}
In order to compute the electric and the thermal conductivities, we consider a perturbation
involving the functions $\{\delta g_{t x_{1}}=U\,\delta H_{t x_{1}},\delta A_{x_{1}}, \delta\chi_1,\delta\sigma_1\}$ which are all taken to be functions of $(t,r)$. Since our background admits a time translation killing vector, we can Fourier decompose our perturbations as
\begin{equation}
f(t,r)=e^{-i \omega v(t,r)}f(r)\,,
\end{equation}
where $v$ is the Eddington-Finkelstein coordinate defined as
\begin{align}
v(t,r)=t+\int_{\infty}^{r}\frac{dy}{U(y)}\,.
\end{align}
 Plugging this ansatz in the equations of motion, we obtain one first order equation and 3 second order ones. We now turn to the boundary conditions for these functions. In the IR we impose in-falling boundary conditions at the horizon which is located at $r=0$
\begin{align}\label{eq:AC_nh_exp}
&\delta H_{t x_{1}}=c_{tx_{1}}+\dots\,,\nonumber\\
&\delta A_{x_1}=c_{x_{1}}+\dots\,,\nonumber\\
&\delta\chi_1=c_{\chi}+\dots\,,\nonumber\\
&\delta\sigma_1=c_{\sigma}+\dots\,,
\end{align}
where $c_{tx_{1}}$ is fixed in terms of the other constants. Thus, we see that the expansion is fixed in terms of three constants, in addition to $\omega$ which we fix by hand since we imagine introducing a time dependent source to the system at that frequency.

For the backgrounds with $\phi_{s}=0$ near the UV \eqref{eq:UVexp}, we have the following expansion
\begin{align}
\label{eq:pertUV}
&\delta H_{t x_{1}}=\delta H_{t x_{1}}^{(s)}+\dots+\frac{\delta H_{t x_{1}}^{(v)}}{r^3}+\dots\,,\nonumber\\
&\delta A_{x_1}=A_{1}^{(s)}+\frac{a_{1}^{(v)}}{r}+\dots\,,\nonumber\\
&\delta\chi_1=\delta\chi^{(v)}+\dots\,,\nonumber\\
&\delta\sigma_1=\delta\sigma^{(s)}+\dots+\frac{\delta\sigma^{(v)}}{r^3}+\dots\,.
\end{align}
together with the constraint
\begin{align}
 12 \omega \delta H_{t x_{1}}^{(v)}=&-\omega \delta H_{t x_{1}}^{(s)}(12 W + 8 i k_s^2 \psi_s^2 \omega-  2 i \omega^3)+  k_s\omega\delta \sigma^{(s)} \psi_s^2(2 k_s^2\psi_s^2 - 2  \omega^2) \nonumber\\
 &-12 i k_s \delta\sigma^{(v)} \psi_s^2 - 4  A_1^{(s)} Q \omega
\end{align}
After naively fixing the sources $\{\delta H_{t x_{1}}^{(s)},A_{1}^{(s)},\delta\sigma^{(s)}\}$ in the UV expansion \eqref{eq:pertUV}, we see that we are left with six constants of integration e.g. $\{c_{x_{1}},c_{\chi},c_{\sigma}, a_{1}^{(v)},\delta\chi^{(v)},\delta\sigma^{(v)} \}$. However, in order to find a solution to our system of ODE's we would need a total of seven constants of integration. That extra constant will come from a particular combination of $\delta H_{t x_{1}}^{(s)}$ and $\delta\sigma^{(s)}$ which can be removed by a boundary coordinate transformation of the form
\begin{align}\label{eq:coord_transf}
x_{1}\to x_{1}+\delta g\, e^{-i\omega t}\,.
\end{align}
Under \eqref{eq:coord_transf}, the sources transform as $\delta H_{t x_{1}}^{(s)}\to \delta H_{t x_{1}}^{(s)}-i\omega \delta g$ and $\delta\sigma^{(s)}\to \delta\sigma^{(s)}+k_{s}\delta g$. Therefore, we see that if instead of naively fixing the sources in terms of the electric field and temperature gradient sources $E$ and $\zeta$, we set
 \begin{align}
 \delta H_{t_{x1}}^{(s)}&=\frac{1}{i\omega}\zeta+i\omega \,\delta g\notag\\
 \delta \sigma^{(s)}&=-k_{s}\,\delta g\notag\\
 \delta A_{1}^{(s)}&=\frac{1}{i\omega} E-\frac{1}{i\omega}\mu\,\zeta
 \end{align}
then we gain the extra constant of integration $\delta g$ we are after, since it can be removed by a coordinate transformation. After performing the above change of coordinates in order to set $\delta g=0$, the boundary electric and heat currents are given by
\begin{align}\label{eq:currents_spont}
\delta J^{\infty}&=\lim_{r\to \infty}\sqrt{-g} \tau F^{x_1\, r} =E -\left(\mu+\frac{i Q}{\omega}\right)\zeta + a_1^{(v)} \,,\nonumber\\
\delta Q^{\infty}& =-T^{x_1}{}_t- \mu\, \delta J^{\infty}= - \mu \, a_1^{(v)}+3 i \frac{k_s\,\psi_s^2}{\omega}\delta\sigma^{(v)}-E\left(\mu+\frac{iQ}{\omega}\right)\nonumber\\
&-i\frac{\zeta}{\omega}\left(3 W+\frac{3}{2}i k_s^2 \psi_s^2\omega-2\mu Q+i \mu^2 \omega\right)+\delta g\, \frac{k_s^2\psi_s^2}{2} \,( k_s^2\psi_s^2- \omega^2)
\end{align}
The transport coefficients can be easily read off after writing
\begin{align}
\delta J^{\infty}&=\sigma\,E+T\alpha\,\zeta\notag\\
\delta Q^{\infty}& =T\bar{\alpha}\,E+T \bar{\kappa}\,\zeta\,.
\end{align}
In our numerics we can then set either $E=0$ or $\zeta=0$ and compute $\delta J^{\infty}$ and $\delta Q^{\infty}$ from \eqref{eq:currents_spont}.

\begin{figure}[h!]
\centering
\includegraphics[width=0.48\linewidth]{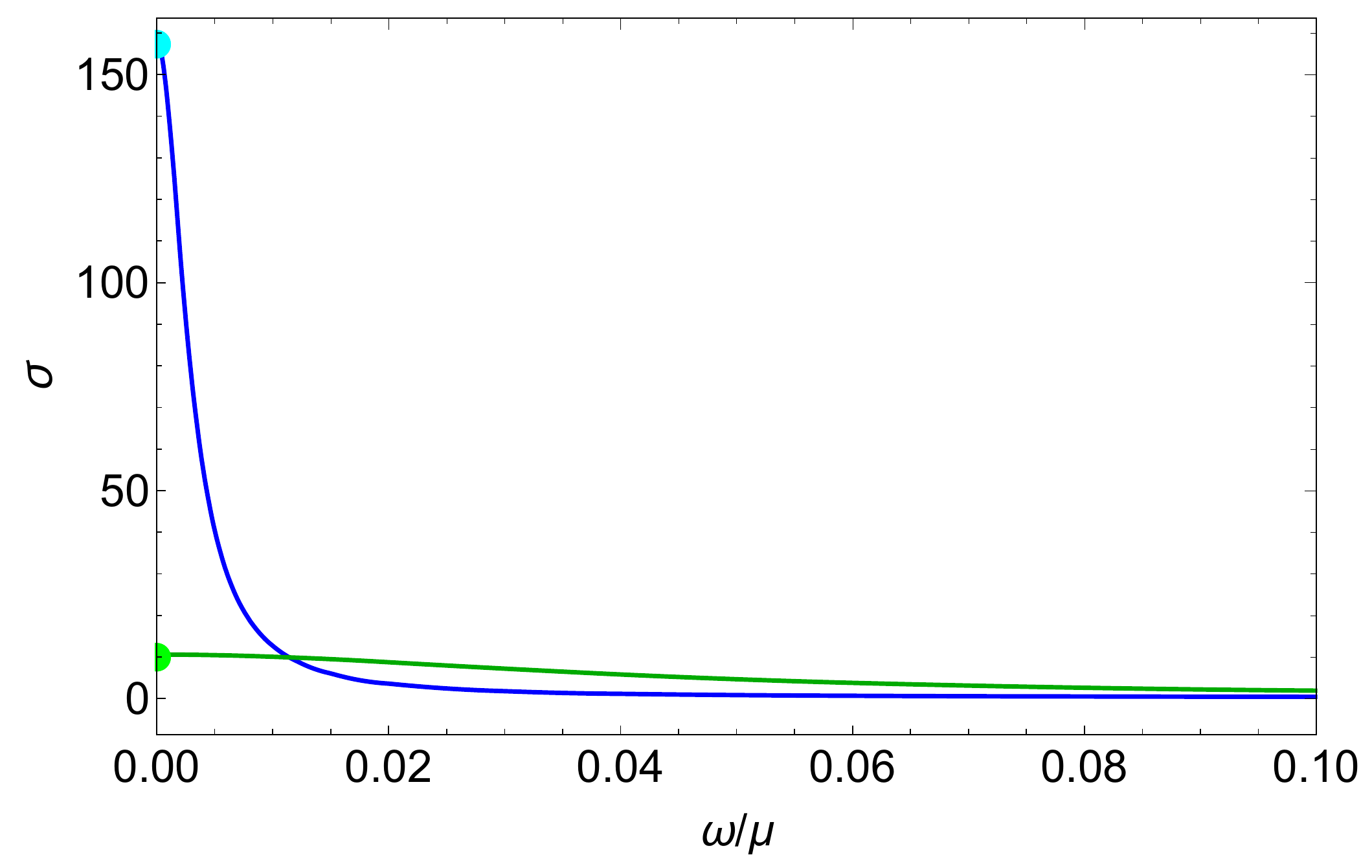}\quad\includegraphics[width=0.48\linewidth]{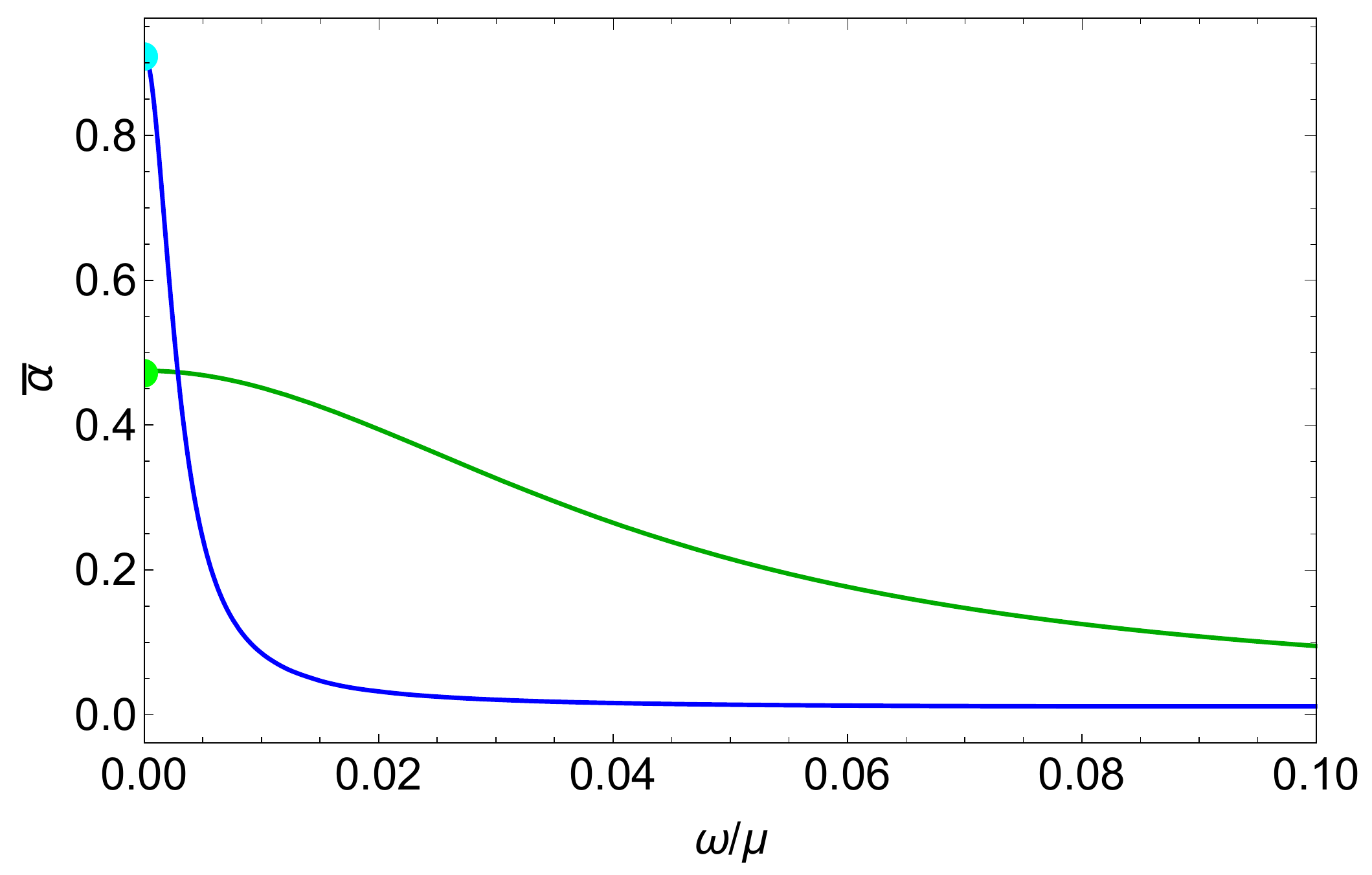}\\
\includegraphics[width=0.48\linewidth]{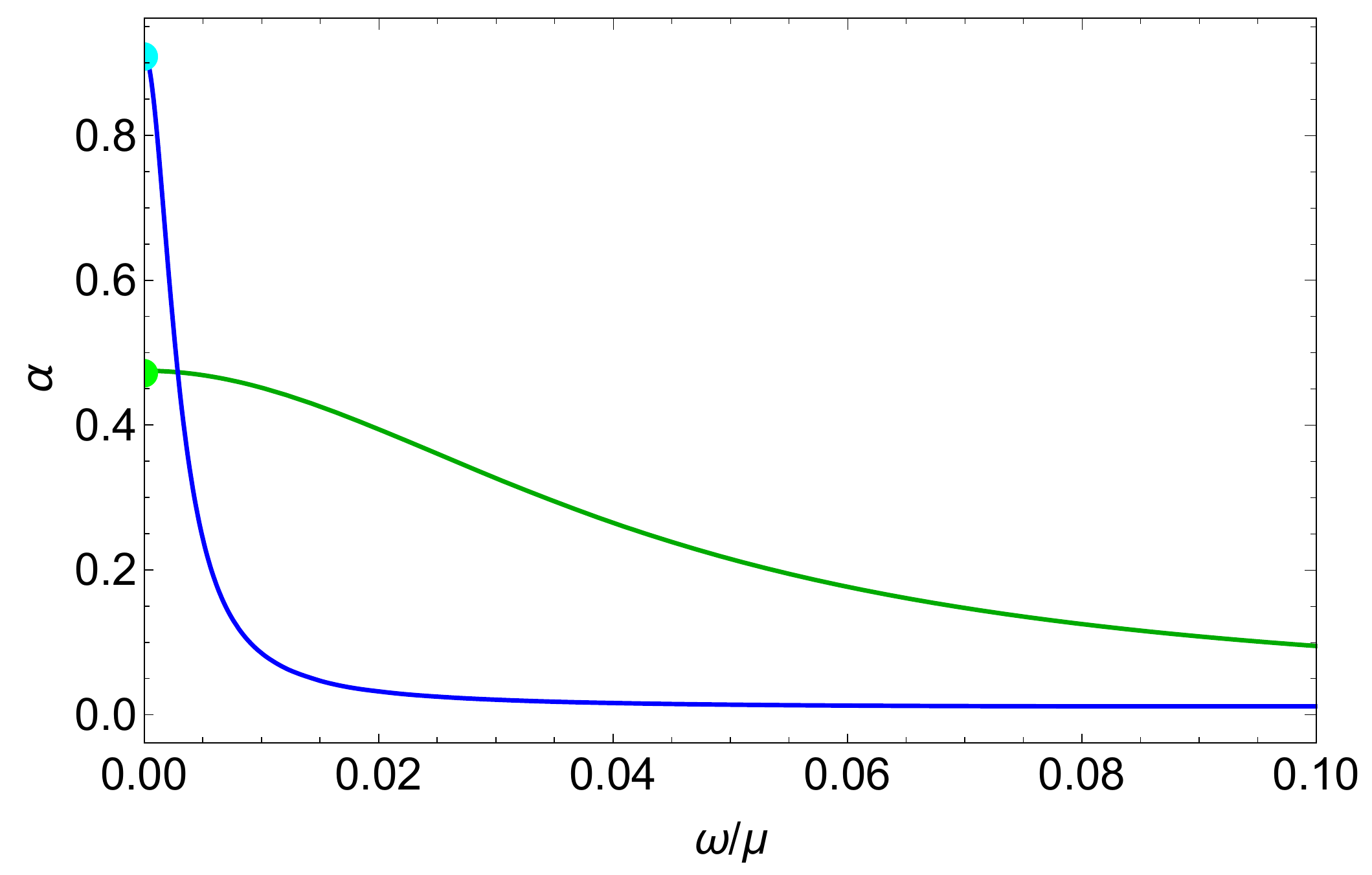}\quad\includegraphics[width=0.48\linewidth]{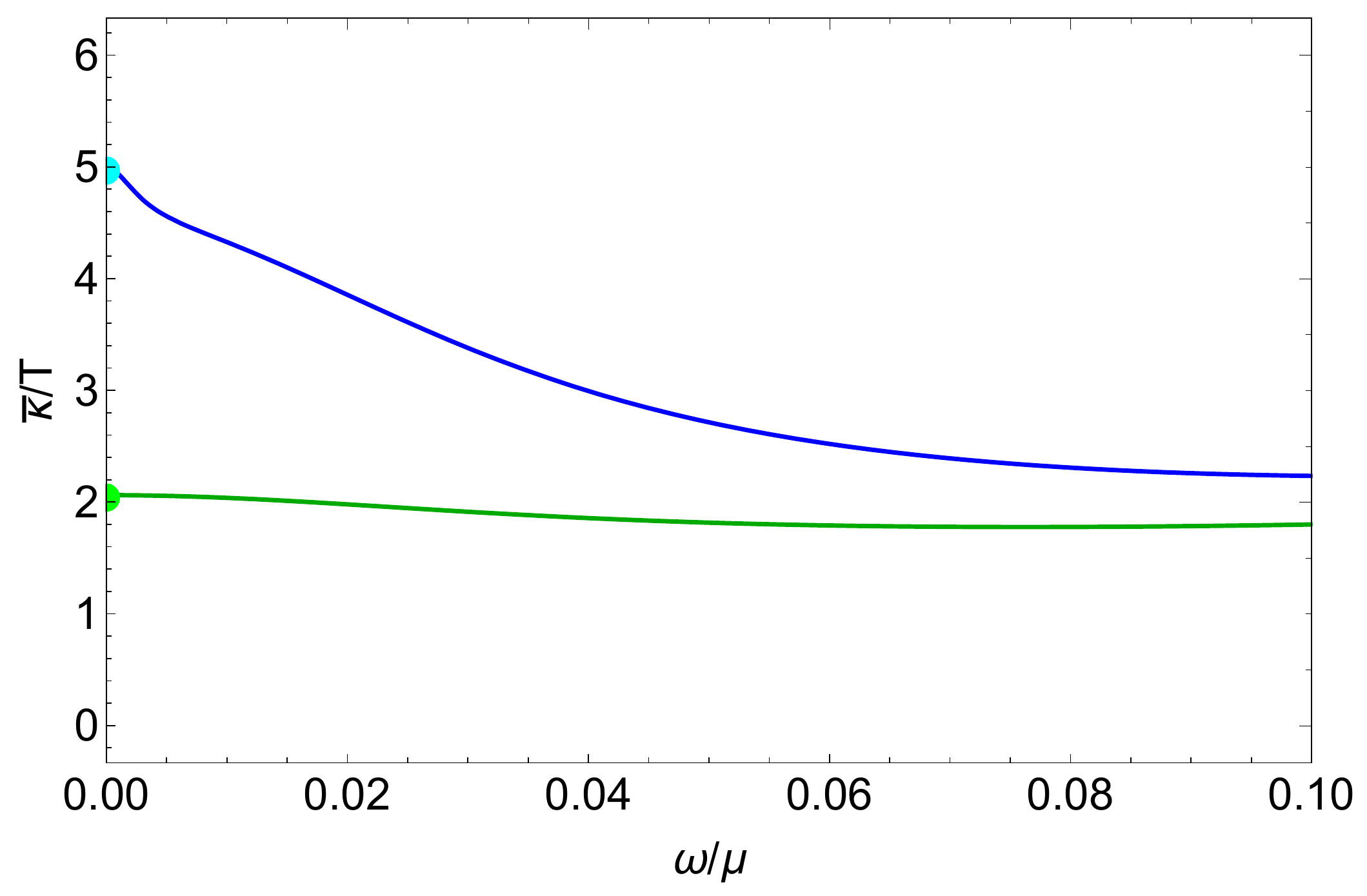}
\caption{Plot of the conductivity matrix as a function of the frequency  for $\phi_{s}=0$ and $T/\mu=1/100$ (blue) and $T/\mu=4/100$ (green). The dots at zero frequency represent the DC conductivities given in \ref{eq:DC_coeffs}.}
\label{fig:AC_cond}
\end{figure}

In figure \ref{fig:AC_cond}, we show the plots of the AC conductivities as functions of the frequency $\omega$ for a background black hole specified by $\{k/\mu,k_s/\mu,\psi_s,\gamma, \delta,T/\mu\}=\{\tfrac{15}{100},\tfrac{3}{10},4,3,\tfrac{1}{2}, \tfrac{1}{100}\}$. The green line corresponds to $T/\mu =4/100$ and the blue to $T/\mu =1/100$. The two dots in cyan and green at zero frequency represent the analytic result \ref{eq:DC_coeffs} for the DC thermoelectric conductivity. From these plots we establish that the DC formulas derived in section \ref{sec:DC}, correctly capture the zero frequency limit of the optical conductivities.

Regardless of the apparently strong momentum relaxation, the electric conductivity in figure \ref{fig:AC_cond} displays the characteristics of a Lorentzian distribution at low frequencies with a large DC value which is very similar to Drude physics. This suggests that the incoherent current gives a large contribution in the ground state of the broken phase. Such a behaviour has been identified before in holographic ground states \cite{Donos:2014uba,Gouteraux:2014hca}, where thermal insulators can have metallic characteristics as far as electric transport is concerned.

\subsection{AC transport with pinning}\label{sec:AC_pinning}
In this section we will carry out the computation for the AC conductivity with the pinning parameter $\phi_{s}$ turned on. The only difference to the subsection \ref{sec:AC_massless} will be the expansion of the perturbation towards the UV part of the geometry. In this case we find that
\begin{align}
\label{eq:pertUVp}
&\delta H_{t x_{1}}^{(s)}=\delta H_{t x_{1}}^{(s)}+\cdots+\frac{\delta H_{t x_{1}}^{(v)}}{r^{3}}+\cdots\,,\nonumber\\
&\delta A_{x_1}=A_{1}^{(s)}+\frac{a_{1}^{(v)}}{r}+\cdots\,,\nonumber\\
&\delta\chi_1=\delta\chi^{(s)}+\dots+\frac{\delta\chi^{(v)}}{r}+\cdots\,,\nonumber\\
&\delta\sigma_1=\delta\sigma^{(s)}+\dots+\frac{\delta\sigma^{(v)}}{r^3}+\cdots\,,
\end{align}
together with the constraint 
\begin{align}
 12 \,\omega& \delta H_{t x_{1}}^{(v)}=-\omega \delta H_{t x_{1}}^{(s)}(12 W +12 S_V \phi_{s}+3 i \omega \phi_{s}^2+ 8 i k_s^2 \psi_s^2 \omega-  2 i \omega^3) -48\, i \,k \,\delta^2 \delta\chi^{(v)} \phi_{s}^2\nonumber\\
 &+ k_s   \omega\delta \sigma^{(s)}\psi_s^2(+3 \phi_{s}^2+2 k_s^2 \psi_s^2 - 2 \omega^2)-12 i k_s \delta\sigma^{(v)} \psi_s^2 - 4  A_1^{(s)} Q \omega+48\, k\,\delta^2 \delta\chi^{(s)}\phi_{s}^2 \omega
\end{align}

This time, the constant term in the expansion of $\delta\chi_{1}$ is part of the source instead of the VEV which is now given by the term falling off like $1/r$.

The near horizon expansion remains as in \eqref{eq:AC_nh_exp}. Consequently the counting argument of constants of integration remains in this case. After naively fixing the constants which determine the boundary sources we would be left with $c_{x_{1}},c_{\chi},c_{\sigma}, a_{1}^{(v)},\delta\chi^{(v)}$ and $\delta\sigma^{(v)}$, which is again one constant less than what we actually need. By performing the boundary change of coordinates \eqref{eq:coord_transf} we see that the sources transform according to
\begin{align}
\delta H_{t x_{1}}^{(s)}&\to \delta H_{t x_{1}}^{(s)}-i\omega \delta g\notag\\
\delta\sigma^{(s)}&\to \delta\sigma^{(s)}+k_{s}\delta g\notag\\
\delta\chi^{(s)}&\to \delta\chi^{(s)}+k\,\delta g\,.
\end{align}
The above suggest that after fixing our sources as
 \begin{align}
 \delta H_{t x_{1}}^{(s)}&=\frac{1}{i\omega}\zeta+i\omega \,\delta g\notag\\
 \delta \sigma^{(s)}&=-k_{s}\,\delta g\notag\\
 \delta \chi^{(s)}&=-k\,\delta g\notag\\
 \delta A_{1}^{(s)}&=\frac{1}{i\omega} E-\frac{1}{i\omega}\mu\,\zeta\,,
 \end{align}
the extra constant $\delta g$ will once again play the role of the integration constant we are after. After performing the change of coordinates in order to set $\delta g=0$, the boundary currents read
\begin{align}\label{eq:currents_exp}
\delta J^{\infty}&=E -\left(\mu+\frac{iQ}{\omega}\right)\zeta + a_1^{(v)}\,,\nonumber\\
\delta Q^{\infty}& =- \mu \, a_1^{(v)}+3 i \frac{k_s\,\psi_s^2}{\omega}\delta\sigma^{(v)}+12 i\delta^2 \frac{k \phi_{s}^2}{\omega} \delta \chi^{(v)}-E\left(\mu+\frac{iQ}{\omega}\right)\nonumber\\
&-i\frac{\zeta}{\omega}\left(3 W+3 S_V \phi_{s} +\frac{3}{2}i k_s^2 \psi_s^2\omega-2\mu Q+i \mu^2 \omega\right)\nonumber\\
&+\frac{1}{4} \delta g\, (48 \,\delta^2\, k^2\,\phi_{s}^2+3 k_s^2\phi_{s}^2\psi_s^2+2 k_s^4\psi_s^4-2 k_s^2\psi_s^2\omega^2)\,,
\end{align}
which reduce to \eqref{eq:currents_spont} after naively setting $\phi_{s}= 0 $.

Having set up the problem for the fully explicitly broken case, we will now study the discontinuity in the DC conductivity when taking the limit $\phi_{s}\to 0$ that we predicted in section \ref{sec:DC}. For this reason we will study the transport at finite frequency as we take the limit of $\phi_{s}\to 0$. This will demonstrate the expected peak in the AC conductivity as a result of the almost gapless mode which couples to the heat current.

\begin{figure}[h!]
\centering
\includegraphics[width=0.48\linewidth]{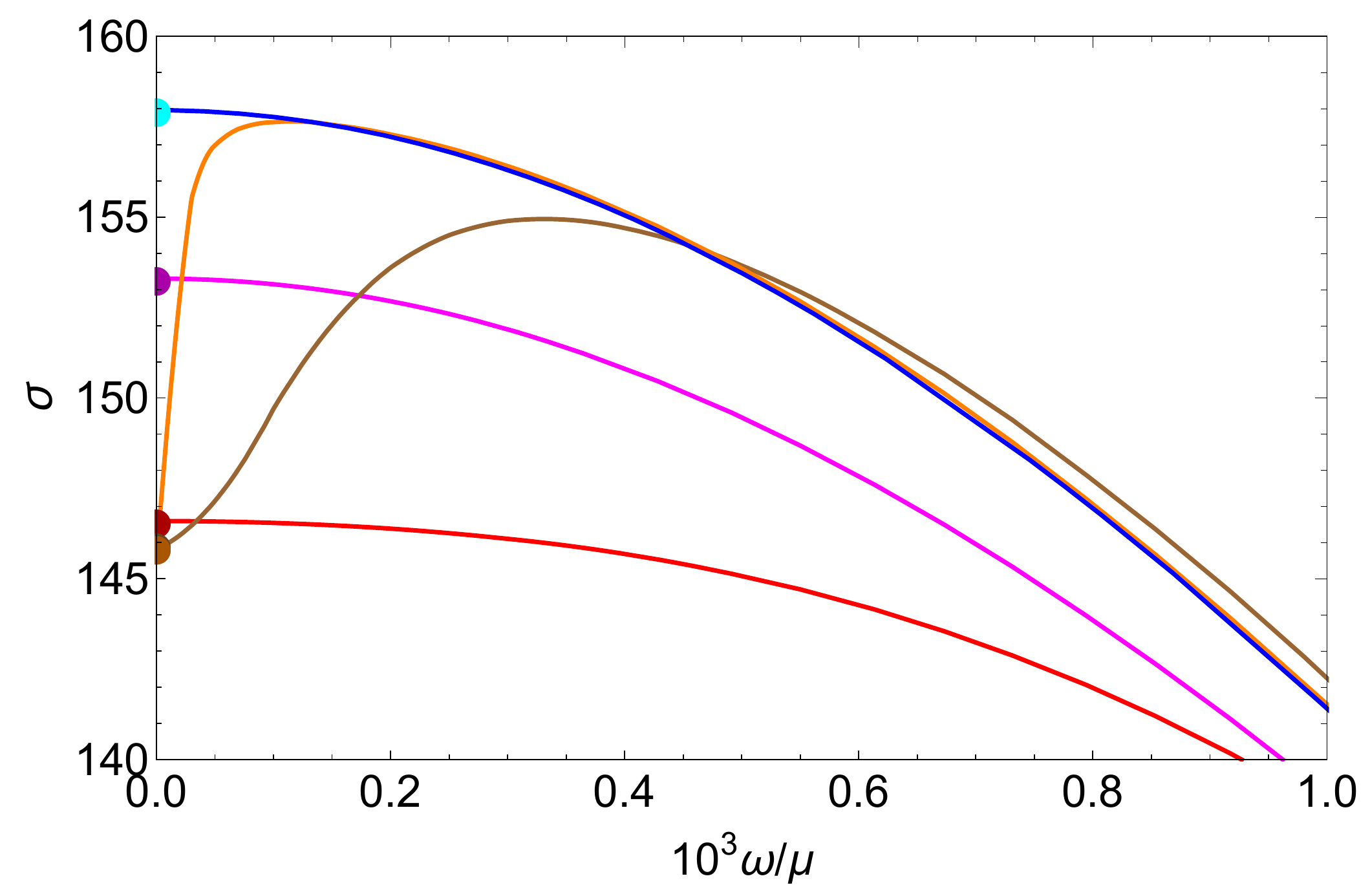} \includegraphics[width=0.48\linewidth]{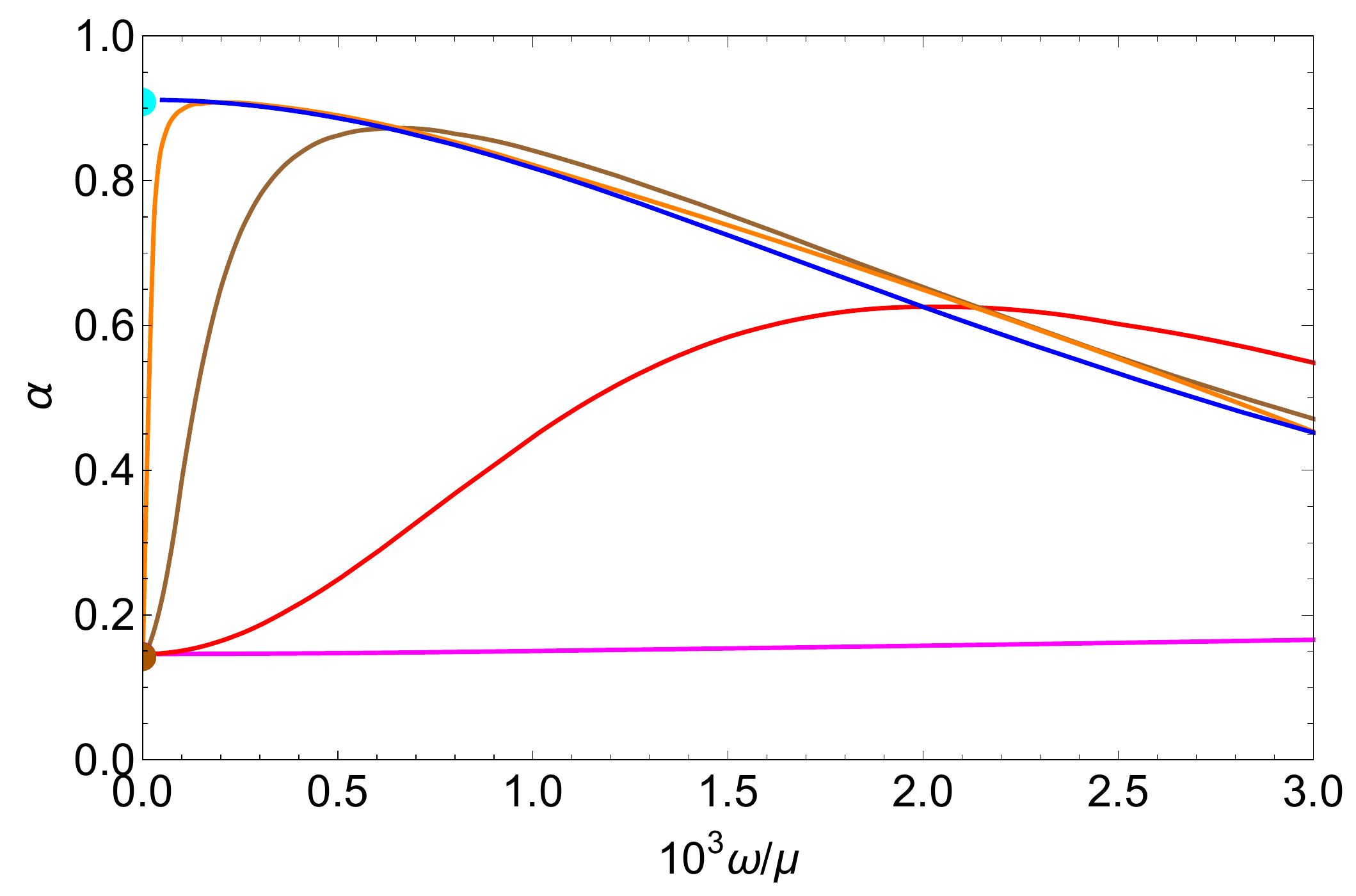}\\
\includegraphics[width=0.48\linewidth]{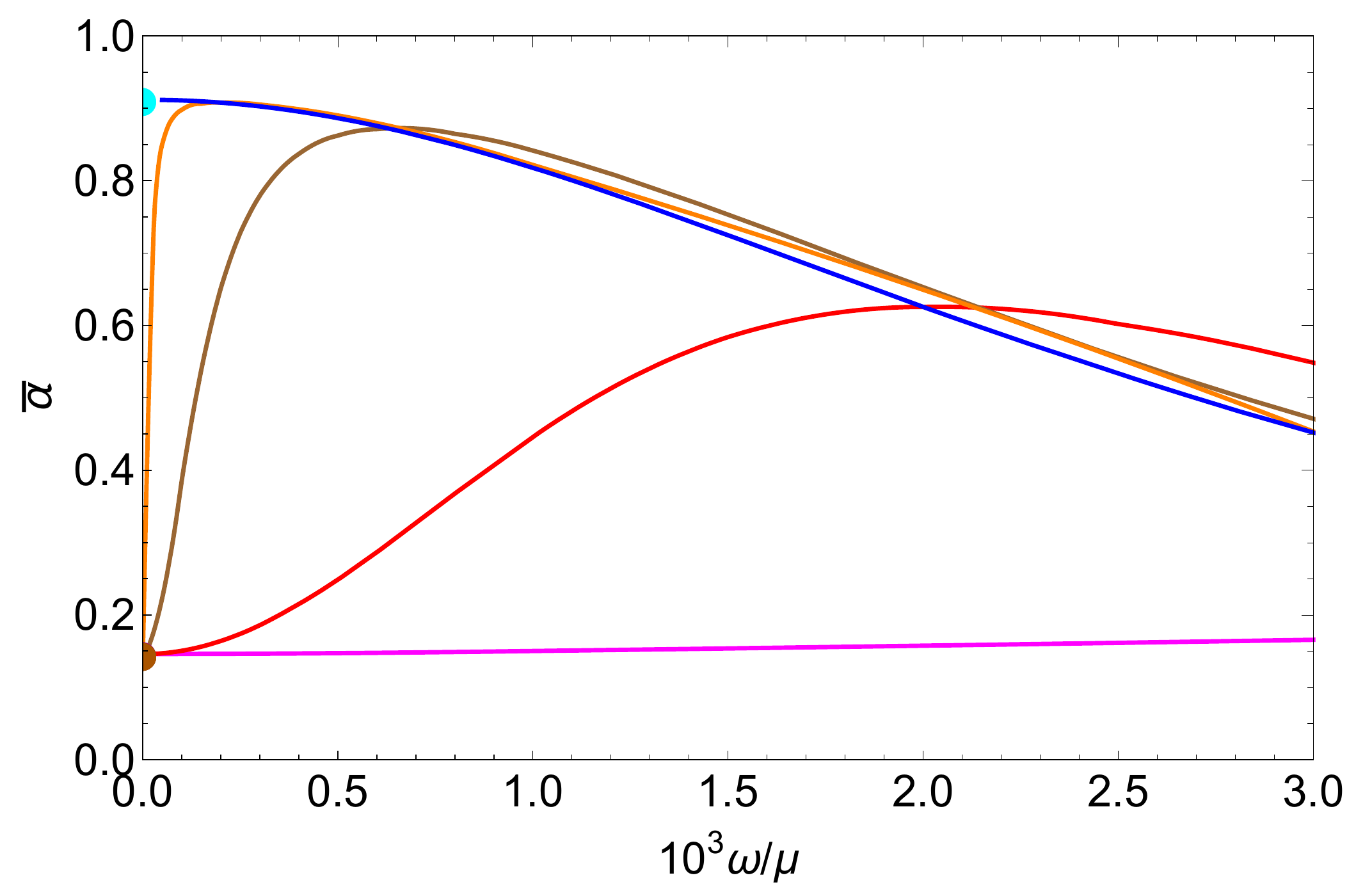} \includegraphics[width=0.48\linewidth]{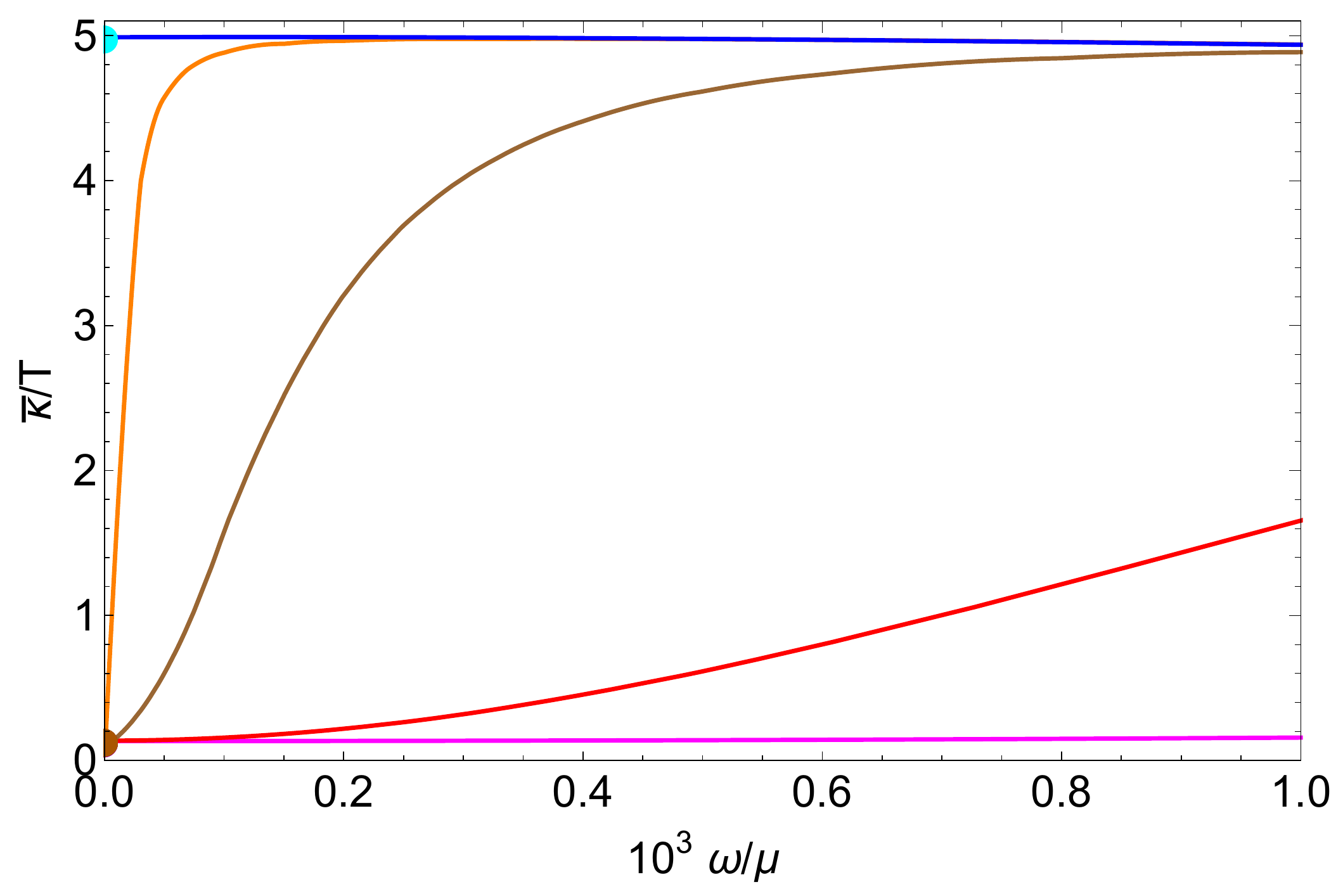}
\caption{Plot of the electric conductivity as a function of the frequency for $T/\mu=1/100$ and for $\phi_{s}=0$ (blue),   $\phi_{s}/\mu=10^{-2}$ (magenta), $\phi_{s}/\mu=10^{-3}$ (red) and $\phi_{s}/\mu=10^{-4}$ (brown) and $\phi_{s}/\mu=10^{-5}$ (orange).}
\label{fig:AC_cond_pinning}
\end{figure}

In particular, we will consider once again the broken phase black holes with $\{k/\mu,k_s/\mu,\psi_s,\gamma, \delta,T/\mu\}=\{\tfrac{15}{100},\tfrac{3}{10},4,3,\tfrac{1}{2}, \tfrac{1}{100}\}$ which we considered also in section \ref{sec:AC_massless}. This time, our aim is to compute the AC transport coefficients after having perturbatively deformed away from $\phi_{s}=0$. The result of this limiting process is shown in figure \ref{fig:AC_cond_pinning} where we demonstrate that the optical conductivity converges to the result of the blue curve of figure \ref{fig:AC_cond} everywhere except for zero frequency. The values in the DC limit are also shown to agree with \eqref{eq:DC_coeffs_pinning} for all non-zero values of $\phi_{s}$.

\section{Discussion and future directions}\label{sec:discussion}

In section \ref{sec:background} we have constructed phases of holographic matter in which a spontaneous density wave coexists with a holographic lattice which breaks translations explicitly. The order parameter is characterised by a wavenumber $k$ which is in general incommensurate to the holographic lattice  associated to wavenumber $k_{s}$. We  note that the wavenumber $k$ is in general fixed by the fact that it has to minimise the free energy. For the particular model in equation \eqref{eq:action}, the thermodynamically preferred branch of black holes has $k=0$ but one can write more complicated models which will have a preferred $k$ away from zero. The main content of the paper is the effect of a finite $k$ density wave on transport which will remain unchanged.

In \cite{Donos:2015gia}, building on \cite{Donos:2014cya}, it was shown that in the DC limit, the thermoelectric transport coefficients are completely fixed after solving for a Stokes flow of a charged fluid which lives on the curved black hole horizon. In section \ref{sec:DC} we significantly generalised the formalism of \cite{Donos:2014cya} in order to properly account for the sliding mode that couples to the heat current. As we saw, the contribution of the gapless mode modifies the naive expression of the DC conductivities in a discontinuous way in the limit where the pinning parameter of the density wave becomes infinitesimally small. In this paper we have used a particular class of models; an interesting question is the generalisation of our treatment to larger classes of holographic lattices and higher derivative theories as in \cite{Donos:2017oym}.

By directly comparing \eqref{eq:DC_coeffs} and \eqref{eq:DC_coeffs_pinning}, we saw that a parametrically small pinning parameter leads to a sudden drop in the DC conductivity. As we demonstrated in section \ref{sec:AC} this effect is accompanied by a transfer of spectral weight from the origin to a frequency which is set by the strength of pinning. An interesting question which we leave to future work is the interplay between temperature and the pinning parameter within the class of models \eqref{eq:action}.

Another direction we are working on \cite{toapp} is the inclusion of a constant magnetic field on the boundary. The resulting Hall angle from the response electric currents is a topic of interest in the literature of the cuprate superconductors. As noted in \cite{Blake:2014yla}, for small magnetic fields, strong explicit breaking of translations leads to a scaling of the Hall angle with temperature  which is in general different from the scaling of the electric conductivity at zero magnetic field. The inclusion of a sliding mode is then a natural question, given the IR nature of the effect observed in \cite{Blake:2014yla}. From a technical point of view, the question would be a natural extension of \cite{Blake:2015hxa,Donos:2015bxe}.

 \section*{Acknowledgements}
 
We would like to thank A. Krikun and T. Andrade for useful discussions. We thank the Lorentz Center for the hospitality during the workshop ``Bringing holography to the lab: Bringing Holography to the Lab: Explaining Strange Metals with Virtual Black Holes'' where parts of this work were finished. AD and CP are supported by STFC grant ST/P000371/1.

%\newpage

\bibliographystyle{utphys}
\bibliography{refs}{}

\end{document}